# Measuring and Monitoring Grid Resource Utilisation

Aleksandar Lazarević[‡], Dr. Lionel Sacks[‡]

[‡]Advanced Communicatios Systems Engineering Group, Dept of Electronic & Electrical Engineering, University College London

**Abstract:** Effective resource utilisation monitoring and highly granular yet adaptive measurements are prerequisites for a more efficient Grid scheduler. We present a suite of measurement applications able to monitor per-process resource utilisation, and a customisable tool for emulating observed utilisation models.

## 1. Introduction

Solving the problems confronting scientists in high energy physics, astronomy, bio-chemistry and social sciences today will require more computational, storage and visualisation resources then ever before. Grid computing[1] attempts to answer those needs by syndicating distributed resources in persistent environments spanning multiple geographical locations and administrative domains. While presenting a unified middleware interface to the user and application developer, computational Grids are built on a heterogeneous mix of hardware and operating systems, with dynamic and conditional availability. Using Grid middleware, high performance clusters can be built using commodity hardware for a fraction of former costs. However, increased complexity of such systems increases running costs through higher administration overheads.

Work herein presented is motivated by the requirement to increase overall utilisation of Grids through more effective job scheduling. Current monitoring tools lack granularity and detail to enable schedulers to make more intelligent decisions[2]. Our view is that Grid computational resource will need to be both time- and space-shared for highest utilisation, and the possibility to monitor and measure resource utilisation on a per-process level will thus be important in control and scheduling. Presented in this paper is a tool developed to simulate various types of Grid application loading patterns, and a suite of monitoring tools able to measure per process resource utilisation. Test results validating the design will be presented, and ways of overcoming detected flaws discussed.

The work is undertaken as part of an EPSRC e-Science sponsored project SO-GRM: Self-Organising Grid Resource Management (GR/S21939) in collaboration with BTExacT.

## 2. GridLoader Application

While more Grids are being deployed as production environments, they often run few applications with similar resource utilisation profiles. As Grids establish themselves as multi-purpose computing platforms, more complex job statistics will develop[3]. GridLoader application was developed out of need for a controllable and tuneable load generator that could be used to test other components of SO-GRM under realistic future utilisation scenarios.

To ensure portability GridLoader was developed in ANSI C without the need for any additional libraries. Based on a state machine, it is able to stress CPU, memory and network subsystems. All aspects of component loading are parameterised either from a configuration file or at run-time using command line options. These parameters give overall bounds and enable coordination of experiments across all the machines in the cluster (e.g. CPU loading time across all machines to follow a Pareto probability distribution function). However, each of the loading modules achieves its target in a probabilistic manner, so that the local monitoring facility will see different loading patterns even for equally parameterised runs.

Current version of GridLoader uses a deterministic state transition table, progressing through network loading, memory allocation and CPU utilisation states in progression. This is similar to an embarrassingly parallel Grid application, such as parameter sweep tools, staging the input data, allocating required memory and proceeding to CPU intensive core calculations that would usually produce a small result data set. A more sophisticated model is possible when GridLoader is used with probabilistic state transition table where all three primary states are entered into many times with changing probabilities.

Reliability of GridLoader was tested by running a set containing 120 jobs taking around 24 hours to complete. Each run of GridLoader was parameterised with the network transfer time, amount of memory to be allocated, CPU load duration and a probability factor influencing the levels of CPU and network loading. The start and end time of each job was recorded with one second resolution, and compared to expected execution times. The job set was first run locally, and then submitted to the same machine using two remote execution scenarios: Secure Shell (SSH) and Globus 2.2. Figure 1. shows the percentage difference in expected versus observed execution times for a sample of 50 jobs.

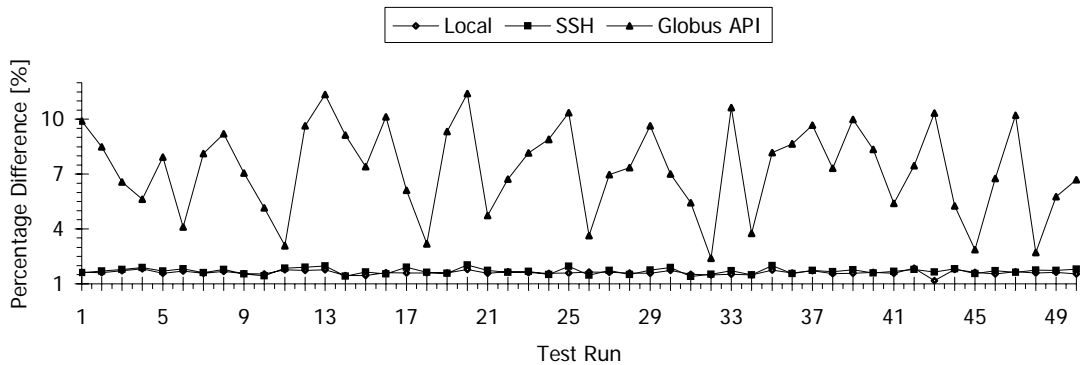

**Figure 1.** Percentage difference in expected and observed execution times

Running on the local node, actual GridLoader execution times are less then 2% greater than expected. This is due to system overhead needed in setting up network transfers and allocating memory, duration of which are not accounted in the timekeeping of the program. As this level of increase in execution time is intrinsic to the operating system, and would be present for all applications, we found that a realistic and accurate simulation of desired load can be achieved with GridLoader.

Since job submission using Globus middleware involves complex authentication and authorization procedure, including creation of proxy X.509 certificate and spawning of other processes, we anticipated that actual execution time will be significantly longer then expected. With the Globus overhead, execution times were increased by 2.5% to 11.5%. Despite using similar authentication model, remote execution through SSH has introduced only a minor overhead.

Recording of start and end times is done outside the GridLoader application, further adding to the observed discrepancy between actual and desired execution time. Systematic errors have also occurred in the rounding of execution times and the resolution of system timers. While these can be further reduced, it would be at a price of more complex and resource demanding code. Bearing in mind that Grid applications run for days and weeks, the possible gain in precision would not bring statistically significant improvement.

## 3. Measurement and Monitoring Suite

The selection of monitoring application for integration into SO-GRM model was driven by scalability, portability and extensibility criteria. When deploying on commodity clusters consisting of thousands of hosts, and with a potential to generate prohibitively large volumes of data, monitoring system must be able to effectively deliver measurement information to the point in the system where it is required. Heterogeneous nature of Grid resources demands an open communication protocol and ability to easily develop custom solution and information providers for specific deployments. Our project has developed a measurement framework based on Ganglia cluster monitoring[4]. Ganglia was selected for its extensible interface, effective storage of data in fixed size round-robin databases, use of XML encoded measurements, and customisable unicast and multicast delivery protocols.

Commonly used schedulers on the Grids today are batch schedulers adapted from legacy cluster systems. Lacking in dynamics and adaptability, they assign single job or process to each CPU in the cluster, ignoring any idle time on that node arising from I/O wait times, code inefficiencies or similar. Equally, current Grid monitoring tools lack granularity and ability to track system resource used by multiple Grid-submitted jobs running on the same machine, leaving the scheduler without vital information if statistical multiplexing of processes is to be used. By creating custom information provider daemons, and integrating with the Ganglia monitoring system, we have reduced the granularity of measurements to recognise jobs, and their child processes, submitted through Grid middleware, determine their resource utilisation, and incorporate that information with the remaining flow of measurement data.

Measurements are collected through either a daemon process running on each node or, depending on site administrator preference, modified Ganglia daemon *gmond*. Frequency of measurements, metrics to be measured (CPU utilisation, memory utilisation etc), and process breakdown (all Grid jobs, per Grid user or application) are customisable. Collected data is encoded in XML, broadcasted via UDP packets, and recorded at one or more monitoring nodes. Round-robin databases used operate as layered wheels storing decreasingly detailed information. By using such hierarchical approach, database size remains constant but contains less detailed data from further in the past. To enable retention of high-frequency data for off-line analysis, one node in the cluster runs a database maintenance process that periodically extracts required metrics and stores them on persistent storage in a format suitable for analysis.

## 4. Experimental Results

Experiments to establish the integration and correct operation of all components were run on SO-GRM's Globus 2.2 test bed. Job set containing 50 jobs and lasting around 8 hours was created with Pareto distributed lengths of network and CPU loading times. All jobs were submitted from one of the machines in the cluster to a different machine in the same cluster using appropriate Globus commands. A simple master-slave scheduling was used, iterating through the job list and allowing 45 seconds between job completion and next job submission for any transient machine loading to settle. Measurement and monitoring suite was configured to record CPU idle percentage and Globus-submitted load percentage with one second resolution, averaged over 15 second intervals.

Figure 2. shows total CPU utilisation and Globus submitted GridLoader CPU utilisation over part of the test run. Data has been adjusted for the impact of monitoring components, although at complete idle they add less than 1% to the total machine load.

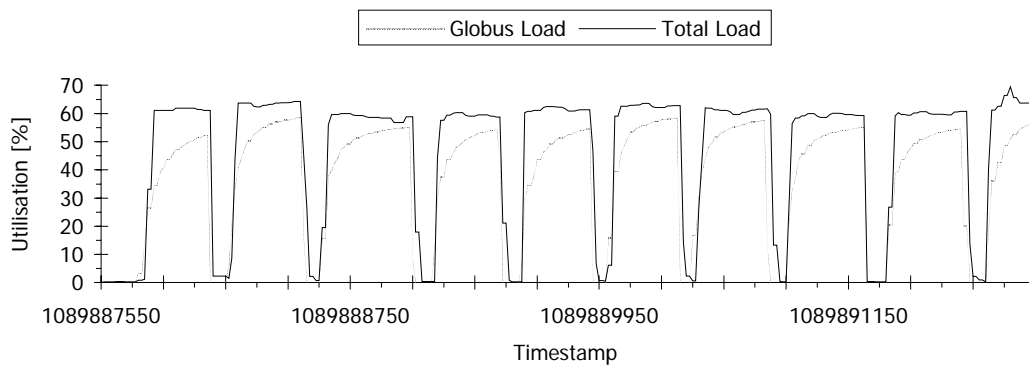

**Figure 2.** Comparison between GridLoader and total CPU utilisation

The measurement capture the difference between GridLoader generated load and total system load which includes various background processes associated with Globus middleware, and kernel time servicing network transfers, memory allocation and process scheduling. Figure 3. (a) shows in more detail percentage difference between total and GridLoader CPU utilisation.

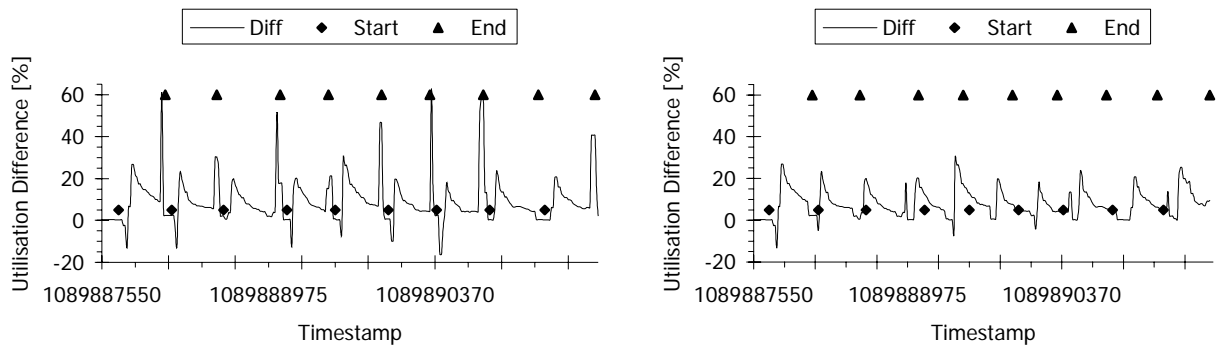

**Figure 3.** (a) Utilisation difference relative to job start and end times; (b) Effect of filtering

The plot in Figure 3. (a) displays discernible and repeated peaking in both positive and negative values. Markers on the same plot indicate recorded job start and end times and there is an obvious correlation between these. These transients occur while the jobs are being staged in and out, and while machine loading is high, but the CPU time is not yet attributed to the process being submitted. To improve the quality of measurement data, samples coinciding with job start and end times are filtered, which leads to a more balanced plot as shown in Figure 3. (b).

Experimental data has exposed a shortcoming in the process monitoring component which leads to a ramp-up effect in the observed loading measurements, as seen in Figure 2. This component uses UNIX standard process reporting which reports CPU usage using a decaying average. Our improved version will use low level measurements of kernel *jiffies* for a more reliable measurement. Resource footprint of the monitoring components was acceptable (below 1%); although an increase was noted as the number of processes grew. This is attributed to the computationally expensive parsing of the processes table required to obtain process IDs of monitored jobs.

## 5. Conclusions

Work presented in this paper forms a stepping stone towards a more effective Grid scheduler. Increased granularity of measurements will enable closer scrutiny of applications running on Grid hosts, help establish the true level of utilisation and provide feedback on inefficient processes. Next generation of schedulers can use this information in predicting resource requirements of applications based on their previous runs and take appropriate actions to ensure quality of service.

The framework is soon to be deployed on UCL Central Computing Cluster consisting of 200 CPUs running Sun Grid Engine. Up to 25 different e-Science projects will use this facility, creating a rich mixture of usage patterns.

Analysis of monitoring data collected from production systems will enable us to parameterise GridLoader and use it for realistic testing of scheduling and management components with more flexibility and variability than it would be possible with trace-replay systems.